\documentclass[twocolumn,english,prb]{revtex4-1}
\usepackage{hyperref}
\usepackage{amsmath}
\usepackage{graphicx}
\usepackage[caption=false]{subfig}

\begin{document}

\title{Thermal transport in disordered one-dimensional spin chains}
\author{Igor Poboiko}
\affiliation{L. D. Landau Institute for Theoretical Physics, 117940 Moscow, Russia}
\affiliation{Moscow Institute for Physics and Technology, Dolgoprudny, Moscow region, Russia}
\author{Mikhail Feigel'man}
\affiliation{L. D. Landau Institute for Theoretical Physics, 117940 Moscow, Russia}
\affiliation{Moscow Institute for Physics and Technology, Dolgoprudny, Moscow region, Russia}
\date{\today}

\begin{abstract}
We study one-dimensional anisotropic XY-Heisenberg spin-$\frac{1}{2}$ chain with weak random fields $h_i^z S^z_i$
by means of Jordan-Wigner transformation to spinless Luttinger liquid with disorder and  bosonization technique.
First we investigate  phase diagram of the system in terms of dimensionless disorder $\gamma = \left<h^2\right>/J^2 \ll 1$ 
and anisotropy parameter $\Delta = J_z/J_{xy}$ and find the range of these parameters where disorder is irrelevant in the
infrared limit and spin-spin correlations are described by  power laws.
Then we use the diagram technique in terms of plasmon excitations to study low-temperature behavior of
heat conductivity $\kappa$ and spin conductivity $\sigma$  in this power-law phase.
The obtained Lorentz number  $L \equiv \kappa/\sigma T $ 
differs from the value derived earlier by means of  memory function method.
We argue also that in the  studied region  inelastic scattering is strong enough 
to suppress quantum interference in the low-temperature limit.
\end{abstract}

\maketitle

\section{Introduction}
One-dimensional disordered spin chain is an excellent example of strongly correlated quantum system that is well suited to study basic properties of such systems. In particular,  studies of disordered spin chains become one of the major playgrounds in the field
of Many Body Localization (MBL) ~\cite{BAA,Mirlin05,HuseReview,mblpalhuse,Serbyn1,Serbyn2,Berkovits14,BarLev14}.
From the experimental viewpoint, quasi-one-dimensional  antiferromagnets~\cite{exp1,exp2,exp3}
attract considerable attention due to very high thermal conductance, that is believed to be related with
 integrability of the clean Heisenberg spin-$\frac12$ chain~\cite{Moore1,Moore2}.
It is known since seminal paper\cite{rggiamarchishulz} that in 1D  competition between interaction and disorder
may lead to delocalization and formation of a ground state that is nearly-free from the effects of disorder, see
also \cite{giamarchibook,bosons2012}.  
Numerical studies~\cite{1dchainNumerical} confirm that qualitative conclusion.
In order to provide delocalization, interaction should be  sufficiently strong  and attractive, so this 
problem bears some resemblance with a model of superconductor-insulator transition in higher-dimensional 
systems~\cite{FIM10}. Looking from that perspective, it seems  useful to develop a quantitative theory of
the delocalized phase of  one-dimensional quantum system with a bare disorder that is "screened" by interactions.
In particular, it is important to study heat transport in such a system, that is expected to be dominated by
the remains of  the disorder potential.

\par Here we will study the properties of anisotropic XXZ spin chain in a random transverse magnetic field, which is described by the Hamiltonian (we assume $J>0$):
\begin{small}
\begin{equation}
\label{eq:H1}
\hat{H} = -J \sum_n \left(\hat{S}_n^x \hat{S}_{n+1}^x + \hat{S}_n^y \hat{S}_{n+1}^y + 
\Delta \hat{S}_n^z \hat{S}_{n+1}^z + \frac{h_n}{J} \hat{S}_n^z \right)
\end{equation}
\end{small}
By means of the Jordan-Wigner (JW) transformation the Hamiltonian (\ref{eq:H1})  can be reduced to the Hamiltonian of interacting spinless fermions (here $\rho_n = c_n^\dagger c_n - \frac{1}{2}$):
\begin{equation}
\label{eq:H2}
\hat{H}=-J\sum_{n}\left(\frac{1}{2}c_{n}^{\dagger}c_{n+1}+h.c + \Delta \rho_n \rho_{n+1} + \frac{h_{n}}{J} \rho_n\right)
\end{equation}
The anisotropy parameter $\Delta$  can be  both positive and negative, which corresponds to the effective attraction or repulsion between JW fermions, respectively.
\par We will consider random fields to be relatively small so that $\gamma = \left<h^2\right>/J^2 \ll 1$, and with zero average $\left<h\right> = 0$. Thus our system is, on average,  symmetric with respect to $z \mapsto -z$ reflection, which translates into the particle-hole symmetry in terms of JW fermions. It ensures that in the quasiparticle spectrum
 $\varepsilon(k)$ only odd powers of $k$ survive.
\par The goal of this paper is to study low-temperature transport properties, spin and heat conductivity,
in the range of parameters $\left(\Delta,\gamma \right)$ where $T=0$ spin-spin correlations decay as a power law
 with a distance. 
The rest of the paper is organized as follows: in Sec.II we study the phase diagram by means of Renormalization Group approach formulated in \cite{rggiamarchishulz}. Sec.III is devoted to formulation and application of the Keldysh approach to the transport properties of disordered Luttinger liquid model that is an appropriate low-energy approximation for the lattice fermion
model (\ref{eq:H2}); in Sec.III A, spin  and heat conductivities ($\sigma$ and $\kappa$) are studied  within the region 
$\frac12 \leq \Delta \leq 1$ where disorder is irrelevant in the RG sense; next, in Sec.III B we discuss specific behavior of $\sigma$ and $\kappa$
 near the critical point $\Delta=\frac12$; the role of quantum interference corrections  and decoherence is discussed in Sec. III C, and the role of spectrum nonlinearity is considered in Sec.III D.
Finally, we present our Conclusions in Sec. IV.

\section{Luttinger liquid description and phase diagram}
In the clean limit $h_n=0$ and in the  region  $-1 < \Delta < 1$  excitation  spectrum 
of the interacting one-dimensional fermion system \eqref{eq:H2} is gapless; then low-energy and long-distance
 properties of the system are known to be described by the Luttinger liquid (LL) model \cite{giamarchibook}.
 It allows to rewrite the  Hamiltonian in terms of fermion density excitations --- plasmons.  LL model is
formulated in terms of canonically conjugated plasmon fields $\left[\phi(x),\Pi(y)\right] = i \delta(x-y)$; in 
the  linear approximation for the quasiparticle spectrum, the Hamiltonian of the LL model reads
\begin{equation}
\label{eq:HLL}
 \hat{H}_{LL} = \frac{1}{2 \pi} \int dx \left(\frac{u}{K} (\partial_x \phi)^2 + u K (\pi \Pi)^2\right).
\end{equation}
Here $u$ is plasmon  velocity and $K$ is dimensionless Luttinger parameter; these parameters are determined,
via the Bethe Ansatz solution for XXZ model,
by the values of $J$ and $\Delta$, see~\cite[p.167]{giamarchibook}:
\begin{equation}
\label{eq:LLparams}
\Delta = \cos \frac{\pi}{2 K}, \quad
u = \frac{J a}{2} \frac{\sin (\pi / 2K)}{1 - 1/2 K} 
\end{equation}
where $a$ is the lattice constant.

\par In our model (\ref{eq:H2}) disorder couples to the fermion density $\rho_n$; in the LL continuum limit it  reads as
$\rho(x) = -\frac{1}{\pi} \partial_x \phi + \frac{1}{\pi a} \cos (2 k_F x - 2 \phi)$. First and second terms in the above expression correspond to the slow ($q \sim 0$) and fast oscillating ($q \sim 2 k_F$) parts. Thus there are two types of scattering of one-dimensional fermions by disorder: 
forward and backward. Forward scattering is  irrelevant within the linear approximation for the  spectrum, since
the corresponding term in the LL Hamiltonian can be  eliminated completely by the redefinition of phase $\phi(x)$.
 Backward fermion scattering with momentum transfer $q \sim 2 k_F$ is the only effect one should take into account then. 
Thus we need only $q \sim 2 k_F$ part of original random potential, this part is described by  the random Gaussian complex field 
$\xi(x)$ with $\left<\xi(x)\xi^*(y)\right> = D \delta(x-y)$ and $D = \left<h^2\right> a$. Disorder contribution to the Hamiltonian reads as follows:
\begin{equation}
\label{eq:Hdis}
\hat{H}_{dis} = -\frac{1}{2 \pi \alpha}\int dx (\xi(x) e^{-2 i \phi} + \xi^*(x) e^{2 i \phi})
\end{equation}

\par Renormalization Group approach to disordered Luttinger liquid was formulated in \cite{rggiamarchishulz}. It is convenient to introduce dimensionless disorder parameter
\begin{equation}
\label{eq:disorder}
g = \frac{2 D a}{\pi u^2} = \frac{8(1-1/2K)^{2}}{\pi\sin^{2}(\pi/2K)}\gamma
\end{equation}
In terms of this parameter and logarithmic scaling parameter $\xi = \ln\frac{\widetilde{a}}{a}$ (with $\widetilde{a}$ being running ultraviolet cutoff), RG equations reads as follows:
\begin{equation}
\label{eq:rgflow}
\begin{cases}
\frac{d u}{d \xi} &= -\frac{u K}{2} g \\
\frac{d K}{d \xi} &= -\frac{K^2}{2} g \\
\frac{d g}{d \xi} &= (3 - 2K) g.
\end{cases}
\end{equation}
\par
\begin{figure}[ht]'
	\includegraphics[width=\columnwidth]{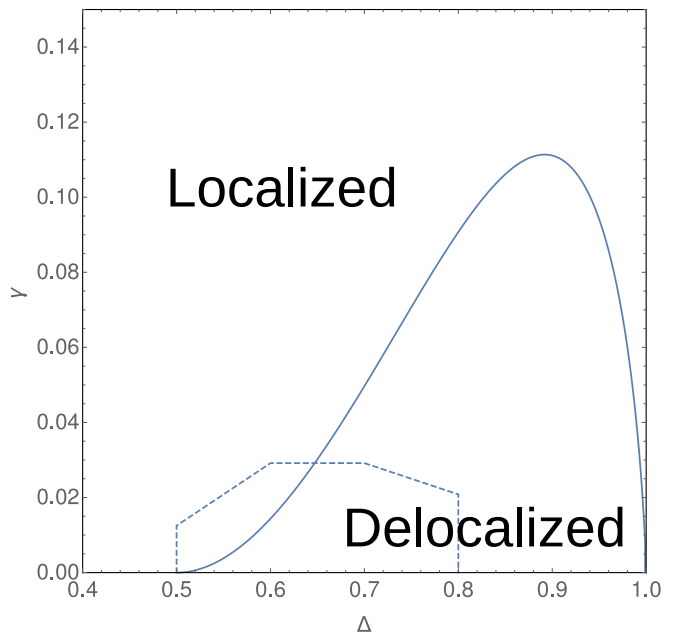}
	\caption{Approximate phase diagram found from the RG calculations for the Luttinger liquid model with linear spectrum. Dashed line corresponds to phase boundary obtained in Ref. \cite{1dchainNumerical}.}
	\label{fig:phasediag}
\end{figure}

 These equations can be solved analytically exploiting its first integral $I(K,g) = \frac{9}{8}(\frac{6}{K} + 4 \ln K - g)$. This solution yields phase diagram shown in Fig. \ref{fig:phasediag}.  
"Delocalized" region lies, in the limit of very weak disorder, in the range $\frac{1}{2} < \Delta < 1$.
Upon increase of $\gamma$, delocalized region shrinks and eventually disappears already at  
$\gamma \approx 0.1$.  Everywhere in the delocalized phase  effective disorder $g(\xi)$  decreases with $\xi$.
Actually the derivation of the RG equations (\ref{eq:rgflow}) as it was performed in~\cite{rggiamarchishulz}
 is valid quantitatively in the vicinity of the point $K = 3/2$ only, where disorder-induced corrections to the 
parameter $K$ are logarithmic; for large $K$  these equations can be used for qualitative analysis only. 
Note that the drop of the critical disorder $\gamma$ near the point $\Delta=1$ is trivially related to the
decrease of the effective Luttinger velocity $u$, see Eqs.\eqref{eq:LLparams}.

Phase diagram obtained by the analysis of RG equations can be compared with the  numeric phase diagram from Ref.\cite{1dchainNumerical} (its boundary is shown by the dashed line in Fig. \ref{fig:phasediag}). According to these numerical data,
the delocalized region covers much smaller part of the phase diagram than the RG calculations predict.
We expect that the major source of this discrepancy is due to inapplicability of the RG equations (\ref{eq:rgflow})
at large $K$ values.  Another reason could be related with the effects of spectrum nonlinearity that becomes important
close to  $\Delta = 1$.  On the other hand,  near the point $K=3/2$  numerical data~\cite{1dchainNumerical}
suggest delocalization at the values of $\gamma$ which are above our critical line; 
we believe that this discrepancy comes from limited  accuracy of the numerical data,
due to finite size effects which becomes most prominent at very weak disorder.

Equal-time spin-spin correlation function $\langle S^+(0)S^-(x)\rangle$ decays as a power law at $\Delta > \frac{1}{2}$,
as one can read of Ref.~\cite{bosons2012} where two-loop RG calculation was performed.  At smaller $\Delta < \frac{1}{2}$ renormalized disorder parameter $g(\xi)$ grows with $\xi$, and one expects exponential decay of $\langle S^+(0)S^-(x)\rangle$
at $ x\geq L_c$, where correlation length $L_c \propto \gamma^{1/(2K-3)}$, see~\cite{rggiamarchishulz}. 

Note that for the case of  XY model with random transverse fields (i.e. $\Delta=0$) exponential decay of $\langle S^+(0)S^-(x)\rangle$ follows directly from single-particle localization in 1D, as proven rigorously in Ref.~\cite{Klein}. 
However, at general $\Delta < \frac{1}{2}$ the relation
between growth of effective disorder upon RG and Anderson localization is far from being obvious, since the RG calculation
~\cite{rggiamarchishulz}  does not contain any multiple-impurity interference effects, see~\cite{mirlintransport,mirlincdw}.


Below we will focus on delocalized phase $\frac{1}{2} < \Delta < 1$, that corresponds to the range of $\frac{3}{2} < K < \infty$, where renormalized disorder constant $g$ is small, and one can obtain transport properties using perturbation theory for bosonic LL model with renormalized parameters.

\section{Transport properties}
\par
\begin{figure}
	\centering
	\subfloat[]{\includegraphics[width=0.45\columnwidth]{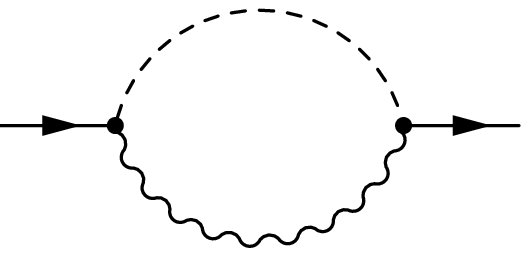}}
	\subfloat[]{\includegraphics[width=0.45\columnwidth]{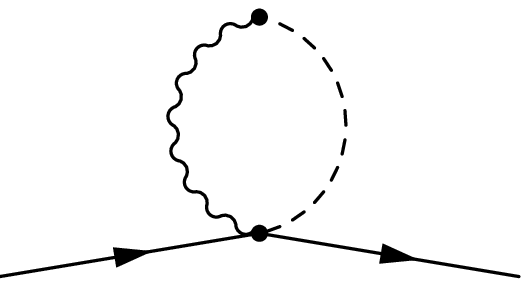}}
	\caption{Lowest order diagrams for the  retarded self energy $\Sigma_R(\omega)$. Dashed lines correspond to disorder average $\left<\xi(x)\xi^*(y)\right>$ and wavy lines correspond to averaging of cosine or sine of boson fields}
	\label{fig:selfenergy}
\end{figure}

 Here we proceed from the Hamiltonian description defined by Eqs.(\ref{eq:HLL},\ref{eq:Hdis}) to the Keldysh action  for the LL model with disorder. Total Keldysh action $S_{tot}$ consists of trivial free boson part $S_0$ and disorder-related part coming directly from Eq.\eqref{eq:Hdis}:
\begin{equation}
S_{dis}= \frac{1}{2\pi\alpha} \int_C dt dx (\xi(x)e^{-2i\phi} + \xi^*(x) e^{2 i \phi})
\end{equation}
We integrate $\exp(iS_{tot})$ over random Gaussian field  $\xi(x)$ and perform Keldysh rotation introducing classical $\phi_{cl} = \frac{1}{2} (\phi_+ + \phi_-)$ and quantum $\phi_q = \phi_+ - \phi_-$ fields components, arriving finally at the effective disorder action
\begin{equation}
S_{dis} = \frac{i D}{\pi^{2} a^2}\int dt_{1}dt_{2} dx\cos 2(\phi_{1cl}-\phi_{2cl}) \sin\phi_{1q}\sin\phi_{2q}
\end{equation}
In order to obtain self-energy for retarded Green function in the lowest order over $S_{dis}$ we consider first order correction to it, which reads
\begin{eqnarray}
i\delta G_{R}^{(1)}(\mathbf{y}-\mathbf{y}^{\prime}) = \frac{iD}{\pi^{2}\alpha^{2}}\intop dt_{1}dt_{2}\int dx_{12} \times \nonumber\\
\times \left\langle \phi_{cl}\phi_{q}^{\prime}\cos(2(\phi_{1cl}-\phi_{2cl}))\sin\phi_{1q}\sin\phi_{2q}\right\rangle_0
\end{eqnarray}
Here we used notation $\mathbf{y} = (y,t)$ and $\phi_i = \phi(y,t_i)$. Performing Wick's contraction, one finds two diagrams (see Fig. \ref{fig:selfenergy}), from which we extract retarded bosonic self-energy $\Sigma_R(\omega)$; the corresponding analytical expression reads
\begin{eqnarray}
\label{eq:selfenergy}
\Sigma_R(\omega) = -\frac{4D}{\pi^{2}a^{2}}\int_{0}^{\infty}dt(1-e^{i\omega t}) \times \nonumber \\
\times e^{2i(G_{K}(t)-G_{K}(0))}\sin2G_{R}(t)
\end{eqnarray}
where bare retarded and Keldysh components of the Green function are as follows
\begin{eqnarray}
G_R^{(0)}(\omega, q) = \frac{\pi u K}{(\omega + i 0)^2 - u^2 q^2} \\
G_R^{(0)}(t,x) = -\frac{\pi K}{2}\theta(t)\theta(ut-|x|) \\
G_K^{(0)}(\omega) = \coth \frac{\beta \omega}{2} (G_R^{(0)}(\omega) - G_A^{(0)}(\omega))
\label{Green0}
\end{eqnarray}
Inverse Fourier transformation of the Keldysh component $G_K^{(0)}(\omega)$ to the real space-time, $G_K^{(0)}(t,x)$,
 is infrared-divergent; it is sufficient to use the difference $G_K^{(0)}(t,x)- G_K^{(0)}(0,0)$ which is finite:
\begin{eqnarray}
 G_K^{(0)}(t,x) - G_K^{(0)}(0,0) = \nonumber \\
 = i \frac{K}{2}\ln\left(\frac{u^{2}\beta^{2}}{\pi^{2}\alpha^{2}}\left|\sinh\frac{\pi(x+ut)}{u\beta}\sinh\frac{\pi(x-ut)}{u\beta}\right|\right)
\end{eqnarray}
At low temperatures $T \ll J$ two different types of contributions to the disorder-induced self-energy can be separated:
virtual transitions with $ T \ll \omega \leq J$  and real (dissipative)  transitions with $\omega \leq T$. First contribution lead to logarithmic renormalization of the model parameters yielding RG equations \eqref{eq:rgflow} described above; 
second contribution yields dissipative behavior of corresponding self-energy $\Sigma_R = -i \omega / u \pi K \tau$. Direct calculation yields the following expression for momentum relaxation rate:
\begin{equation}
\frac{1}{\tau(T)}=\frac{2 D K}{u} \frac{\Gamma^{2}(K)}{\Gamma(2K)}\left(\frac{2\pi a T}{u}\right)^{2K-2}
\label{tau}
\end{equation}
According to Eq.(\ref{tau}),  product $T\tau \propto T^{3-2K}$  diverges as $T \to 0$ in the
delocalized phase.
Full Green function then reads as follows:
\begin{equation}
G_R(q,\omega) = \frac{\pi u K}{\omega^2 - u^2 q^2 + i \omega / \tau}.
\end{equation}

\subsection{Spin and heat conductivities}
To obtain transport properties, one can apply Kubo formulas. Expressions for spin and energy currents can be derived from corresponding continuity equation $\partial_t \rho_\alpha + \nabla j_\alpha = 0$ (index $\alpha$ corresponds to either spin or energy), and using classical equations of motion. For the Hamiltonian of the form $\hat{H}=\int dx\rho_{E}(\phi(x),\nabla\phi(x),\Pi(x))$, equations of motion reads as follows:
\begin{equation}
\partial_{t}\phi=\frac{\partial\rho_{E}}{\partial\Pi},\quad\partial_{t}\Pi=-\frac{\partial\rho_{E}}{\partial\phi}+\nabla\frac{\partial\rho_{E}}{\partial(\nabla\phi)}
\end{equation}
so energy density obeys the following continuity equation:
\begin{equation}
\partial_{t}\rho_{E}=\frac{\partial\rho_{E}}{\partial\phi}\partial_{t}\phi+\frac{\partial\rho_{E}}{\partial\nabla\phi}\partial_{t}\nabla\phi+\frac{\partial\rho_{E}}{\partial\Pi}\partial_{t}\Pi=\nabla\left(\frac{\partial\rho_{E}}{\partial\nabla\phi}\frac{\partial\rho_{E}}{\partial\Pi}\right)
\end{equation}
and similarly for spin density. Considering total Hamiltonian consisting of two contributions \eqref{eq:HLL} and \eqref{eq:Hdis}, we arrive at the following expressions for currents:
\begin{equation}
j_{s} =\frac{1}{\pi}\partial_{t}\phi,\qquad 
j_{E} =-\frac{u}{\pi K}\partial_{t}\phi\nabla\phi.
\label{currents}
\end{equation}
We  emphasize  that Eqs.\eqref{currents} provide exact (within Luttinger liquid approximation) expressions 
for both spin and thermal currents.  Surprisingly, in the LL approximation the energy current does not contain any terms related to the presence of backscattering. In Appendix~\ref{AppendixA} we provide a detailed derivation of the energy 
current, starting from the lattice fermion model~\eqref{eq:H1}, and show that  backscattering does produce additional terms
for the energy current, but these terms vanish in the continuous LL limit, when $a \to 0$ at some fixed value of the
 product $Ja$.

\par Spin transport is governed by the single-plasmon Green function, while for energy transport we need to calculate
 correlation function of four $\phi$ fields. Applying Kubo formula for spin conductivity, we  reproduce Drude-like
 result of Refs.\cite{mirlincdw,orignac}.
\begin{equation}
\label{eq:spincond}
\sigma(\omega) = \frac{i\omega}{\pi^{2}}G_{R}(\omega,q=0) = \frac{uK}{\pi}\frac{\tau}{1 -i\omega\tau }
\end{equation}
valid at $\omega \ll T$.
\par Thermal conductivity $\kappa$ is expressed in terms of so-called ``thermal susceptibility'' $\chi_E(q,\omega)$ as $\kappa(\omega) = -\frac{i \beta}{\omega + i 0} (\chi_E(0,\omega) - \chi_E(0, 0))$. Introducing short notation $\textbf{x} = (x,t)$, and $\textbf{q} = (q,\omega)$, expression for thermal susceptibility in real space reads as follows:
\begin{equation}
\chi_{E}(\textbf{x}_1-\textbf{x}_2) = i\frac{u^{2}}{\pi^{2}K^{2}}\left\langle (\partial_{t}\phi_{1}\nabla\phi_{1})_{cl}(\partial_{t}\phi_{2}\nabla\phi_{2})_{q}\right\rangle 
\end{equation}
and $\chi(\textbf{q})$ is the Fourier transform of this expression.
\begin{figure}
	\centering
	\includegraphics[width=0.5\columnwidth]{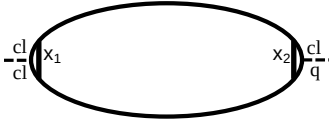}
	\caption{Loop diagram for ``thermal susceptibility'' $\chi_E$. Thermal current vertices act as the following combinations of derivatives: $(\partial_t \nabla^\prime + \partial^\prime_t \nabla)$, with derivatives $\partial_t$, $\nabla$ acting on one $\phi$ field in the vertex and $\partial_t^\prime$, $\nabla^\prime$ acting on another one.}
	\label{fig:loop}
\end{figure}

In the \textit{dc} limit $\omega \to 0$ one finds 
$\kappa = -i\beta\left.\frac{\partial\chi_E}{\partial\omega}\right|_{\omega=0}$. Applying Wick's theorem, one finds:
\begin{eqnarray}
\chi_E(\mathbf{q}^\prime) & = -i\frac{u^{2}}{2\pi^{2}K^{2}}\int(d^2 \mathbf{q})\left[2\omega q-\frac{\omega^{\prime}q^{\prime}}{2}\right]^{2} \times \nonumber \\ 
& \times G_{K}(\mathbf{q}+\frac{\mathbf{q}^{\prime}}{2})G_{A}(\mathbf{q}-\frac{\mathbf{q}^{\prime}}{2})
\end{eqnarray}
Calculating it in the $\omega \to 0$ limit  with ``dressed'' Green functions, we arrive at:
\begin{equation}
\label{eq:thermalcond}
\kappa=\frac{1}{4}u\beta^{2}\tau\int\frac{d\omega}{2\pi}\frac{\omega^{2}}{\sinh^{2}\frac{\beta\omega}{2}}=\frac{\pi}{3}uT\tau
\end{equation}
\par Comparison between Eqs.(\ref{eq:spincond},\ref{eq:thermalcond}) provides us with the value of 
the Lorentz number 
\begin{equation}
\label{Lorentz}
L = \frac{\kappa}{\sigma T} = \frac{\pi^2}{3 K}
\end{equation}
which matches its standard Fermi liquid value $L_{FL} = \frac{\pi^2}{3}$ for $K = 1$. 
 Note that our result \eqref{Lorentz} differs from one obtained in \cite{orignac} by means of memory function formalism.
We believe that this discrepancy is due to limitations of the memory function formalism ~\cite{orignac} which is based on 
extrapolation from the large-$\omega$ region to the static limit. Indeed, frequency-dependent
thermal conductance $\kappa(\omega)$  depends on two different frequency scales, $T$ and $1/\tau$; according to
Eq.(\ref{tau}), in the region $K > 3/2$ one always has $T \gg 1/\tau(T)$ in the low-temperature limit.
In order to obtain static thermal conductivity, one should be able to compute
 $\kappa(\omega)$ at $\omega\tau \ll 1$, whereas memory function method is based upon the calculation of the
high-frequency limit $\kappa (\omega \geq 1/T)$ and further extrapolation to zero frequency. 
We believe that the presence of two parametrically different frequency scales $1/\tau$ and $T$ 
 makes such an extrapolation unreliable.

\par The above  calculation leading to Eqs.\eqref{eq:thermalcond} and \eqref{Lorentz}  should be performed, in general,
 with the parameters $(g,K)$ renormalized (due to RG equations \eqref{eq:rgflow}) down to the  temperature scale $\xi_T$.
 If bare parameters $(g,K)$ are in the bulk of delocalized phase (not too close to the  transition line) 
 one can neglect renormalization of $K$ and $u$ due to disorder, leading to the results for spin and thermal conductivities  
which depend on the scale $\xi_T$ via scattering time $\tau$ only, see~\eqref{tau}. Then the result
is given by Eqs. \eqref{eq:spincond}, \eqref{eq:thermalcond} with bare parameters.
\par Near the transition line one should take renormalization of all the parameters simultaneously. 
Below we will see how it affects physical properties of the system.

\subsection{Vicinity of the point $\Delta = \frac{1}{2}$ }
Expanding first integral of system \eqref{eq:rgflow} by $K-\frac{3}{2}$, or, equivalently,  $\Delta - \frac{1}{2}$, one obtains:
\begin{equation}
I - I_{c} = 
\frac{27}{\pi^{2}}\left(\Delta-\frac{1}{2}\right)^{2}-\frac{16}{3 \pi}\gamma
\end{equation}
The equality $I=I_c$  yields the phase boundary of the delocalized state in the form 
$\left(\Delta-\frac{1}{2}\right)^{2}=\frac{16\pi}{81} \gamma $. 

\par Solution of the equations \eqref{eq:rgflow} can be expressed in terms of vicinity to transition line $\alpha = \sqrt{I - I_c} \ll 1$:
\begin{equation}
\begin{cases}
u(\xi) &= u \exp\left(\frac{2}{3} K(\xi)-1\right) \\
K(\xi) &= \alpha \coth \alpha (\xi + \xi_0) \\
g(\xi) &= \frac{8 \alpha^2}{9 \sinh^2 \alpha (\xi + \xi_0)} 
\end{cases}
\end{equation}
where $\xi_0$ depends on initial values of parameters. Considering temperature to be low enough (so $\xi_T \gg |\xi_0|, 1/\alpha$), one obtains low-temperature behavior of renormalized parameters 
 $g(\xi_T) \simeq \alpha^2 \exp(-2 \alpha \xi) \simeq \alpha^2 (T/J)^{2 \alpha}$ and $K(\xi_T) - \frac{3}{2} \simeq \alpha = \mathrm{const}$. 
 
\par  Now we  repeat the above calculations leading to nonzero $\mathrm{Im}  \Sigma(\omega)$ and obtain Drude-type formulae
 with corrected power-law exponent $\alpha$:
\begin{equation}
\label{eq:modifiedpowerlaw}
 \sigma \simeq \alpha^{-2} a (T / J)^{-1-2\alpha} \, \qquad \kappa \simeq J \alpha^{-2} a (T/J)^{-2 \alpha}
\end{equation}
The Lorentz number is still given by Eq.(\ref{Lorentz}) once renormalization $K \to K(\xi_T)$ is taken into account.
 Modifications of $K$ and $\alpha$  are negligible if $g \ll (K-\frac{3}{2})^2$.

\subsection{Smallness of the interference corrections.}

Our result for the heat conductance, Eq. (\ref{eq:thermalcond}),  was obtained within Drude-type approximation.
Since our system is one-dimensional, some care should be exercised to check if the effects of  quantum interference and
Anderson localization could affect that result.  To begin with, it is useful to employ the result of
 Ref.\cite{mirlintransport} where the same issue was considered for  disordered  Luttinger liquid with a 
weak interaction, $|K-1| \ll 1$. Namely, it was found in~\cite{mirlintransport} that  interference corrections are negligible at sufficiently high temperatures $ T \geq \tau^{-1}(T)(K-1)^{-2}$.  We are working at $K > 3/2$ and the corresponding 
condition is just $T \gg 1/\tau(T)$  which is always fulfilled at low temperatures according to Eq.(\ref{tau}). 

\begin{figure}
	\includegraphics[width=0.4\columnwidth]{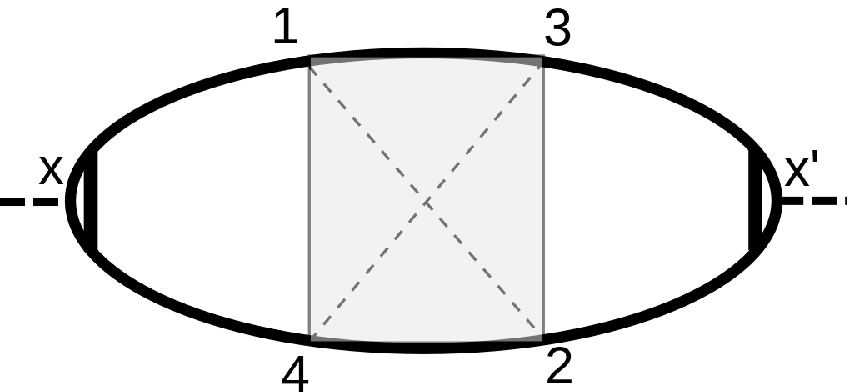}
	\includegraphics[width=0.4\columnwidth]{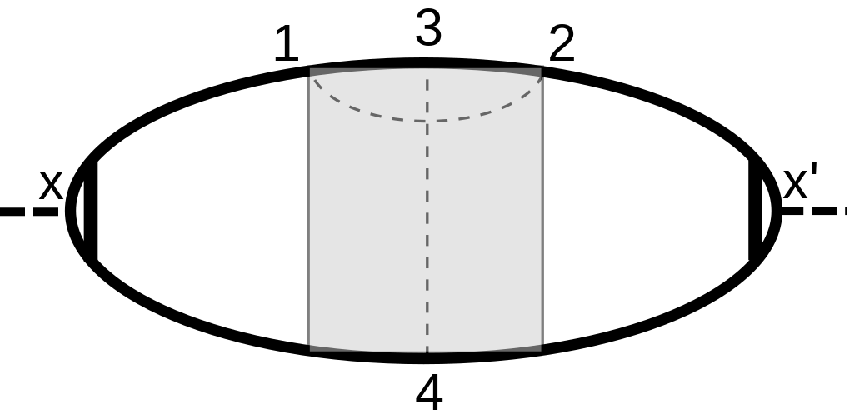}
	\caption{Non-trivial corrections to ``thermal susceptibility''. Dashed lines correspond to same impurity, and grayed area correspond to average of cosine and sine of $\phi$ fields. See main text for the analytical expressions.}
	\label{fig:loopcorr}
\end{figure}

\begin{figure}
	\includegraphics[width=0.5\columnwidth]{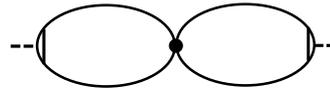}
	\caption{Effective form of the diagram for the vertex correction to thermal susceptibility,  valid
in the  leading order of expansion over  $1 / T \tau(T) \ll 1$ }
	\label{fig:loopeffcorr}
\end{figure}

To estimate interference corrections more accurately, we examine expression for ``thermal susceptibility'' to higher order in $S_{dis}$ adding impurity lines connecting upper and lower Green functions drown in Fig.3.  First order correction (with single impurity line) vanishes at zero external momentum due to gradient structure of energy current vertex. First non-trivial corrections are due to diagrams shown in Fig. \ref{fig:loopcorr}; the corresponding analytical expressions yield:
\begin{eqnarray}
\delta\chi_{E}(\mathbf{x}-\mathbf{x}^{\prime})=\frac{1}{2}\left(\frac{iD}{\pi^{2}\alpha^{2}}\right)^{2}\int dt_{1}dt_{2}dt_{3}dt_{4} \times \nonumber \\
\times \int dx_{12}dx_{34} \Bigg\langle\partial_{t}\phi_{cl}(\mathbf{x})\nabla\phi_{cl}(\mathbf{x}) \times \nonumber \\
\times \left(\partial_{t}\phi_{cl}(\mathbf{x}^{\prime})\nabla\phi_{q}(\mathbf{x}^{\prime})+\partial_{t}\phi_{q}(\mathbf{x}^{\prime})\nabla\phi_{cl}(\mathbf{x}^{\prime})\right)\times \nonumber\\
\times\cos(2\phi_{1cl}-2\phi_{2cl})\sin\phi_{1q}\sin\phi_{2q}\times \nonumber\\
\times\cos(2\phi_{3cl}-2\phi_{4cl})\sin\phi_{3q}\sin\phi_{4q}\Bigg\rangle
\end{eqnarray}
Grayed box correspond to sine and cosine average and consists of infinite number of boson propagators connecting all the points. Generally speaking, such box depends on all the ingoing energies and momentums. However, direct calculation shows that it contains factors $e^{2 i (G_K(t_i - t_j) - G_K(0))} \propto 1 / \sinh^{2 K} \pi T (t_i - t_j)$, which impose effective  constraint for time differences: any such diagram is very small unless the condition $|t_i - t_j| \leq 1/T$ is
fulfilled. On the other hand,  typical time scale for the dressed ``external'' (w.r.t. to the "grey area")
 propagators is $\tau(T) \gg 1/T$; therefore, up to the leading order in  $1 / T \tau(T) \ll 1$
 one can try to shrink all four space-time impurity points in Fig. \ref{fig:loopcorr} into single one (see Fig. \ref{fig:loopeffcorr}). However, calculation of the remaining integrals result in a zero result, due to
  vector structure of the current vertex.  Therefore, nonzero vertex corrections appear in the next order
in $1 / T \tau(T) \ll 1$ only,  and are  small at low $T$ in the whole ``delocalized'' phase $K > 3/2$.


\subsection{Spectrum nonlinearity effects}
At the Heisenberg isotropic point $\Delta = 1$ in the clean system the spectrum of excitations is quadratic and system is no longer described by Luttinger liquid model. In the vicinity to this point plasmon velocity vanishes as $u = J a \sqrt{(1 - \Delta)/2}$; since dimensionless disorder strength $g$ depend on velocity $u$ and interaction parameter $K$, see Eq.\eqref{eq:disorder}, this narrows the region where perturbation theory  in powers of small $g$ is applicable to
$\left<h^2\right> / J^2 \ll (1-\Delta)^{3/2}$.

\par However spectrum nonlinearity effects at finite temperatures might become relevant long before $\Delta = 1$ critical point. Let us make some estimates. Due to particle-hole symmetry, only odd powers in quasiparticle spectrum survive; first non-vanishing contribution to dispersion relation will be $\delta \epsilon \sim \frac{u}{a} (k a)^3$. At finite temperatures this yield new energy scale $\delta \epsilon \sim T (T a / u)^2$; such energy scale should be compared with scattering ratio $1/\tau \sim u g / a (T/J)^{2 K - 2}$. Therefore we conclude that spectrum nonlinearity will be important and should be taken into account when $\left(T/J\right)^{5-2K}\geq\left\langle h^{2}\right\rangle /J^{2}$.
\par For $K < 5/2$ it leads to the  threshold for the temperature, above which nonlinearity effects are expected to be important, $T_* \sim J (\left<h^2\right>/J^2)^{1/(5K-2)}$; on the contrary, at $K > 5/2$  nonlinearity is always  important at low temperatures. In terms of the $\Delta$ parameter, the borderline at $K=5/2$  corresponds to $\Delta = \cos \pi/5 = (1 + \sqrt{5})/4 \approx 0.81$.

\section{Conclusions}

We have analyzed spin and thermal conductance of XXZ spin chain with random-field disorder
 in the parameter region where major source of
disorder (backscattering of Jordan-Wigner fermions) is suppressed by quantum fluctuations and irrelevant in the RG sense
at low temperatures.  Within the standard bosonization scheme the problem is reduced to the Luttinger liquid model with
linear spectrum  $\varepsilon(k) \simeq u k$ 
and Luttinger interaction parameter $K$  in the range $3/2 < K < \infty$, which corresponds to
$1/2 < \Delta < 1$ in terms of original anisotropy parameter $\Delta = J_z/J$ of the spin chain.
We derive a phase boundary in terms of $\Delta$ and normalized disorder $\left<h^2\right> / J^2 $ and
compare it with the numerical result of Ref.~\cite{1dchainNumerical}.
Then we use diagrammatic Drude-like calculation of thermal and spin conductivities and found Lorentz number,
\eqref{Lorentz}, different from the previous result~\cite{orignac}.  We also argue that quantum interference
(the effects beyond Drude approximation) is irrelevant at low temperatures  due to strong enough 
inelastic scattering at $1/2 < \Delta < 1$.

These results were obtained neglecting forward scattering of Jordan-Wigner fermions by disorder,
 which is allowed as long as the approximation of LL model with linear spectrum is employed.
However, this approximation is not evidently correct everywhere in the delocalized phase. 
We estimated region  where it might lead to qualitatively different low-temperature behavior 
as $ \Delta > \cos \frac{\pi}{5} $. The effects of spectrum non-linearity will be considered in the separate publication.

We have not studied the region $\Delta < \frac12$ where localization due to disorder is expected;
here it is very interesting to consider the close vicinity of the transition point,  $\frac12 - \Delta \ll 1$ and
$\left<h^2\right> / J^2 \ll 1$ and search for the existence of localization-delocalization threshold as function of
excitation energy, like the one studied in~\cite{FIM10,FI15} for the Bethe lattice model.

This research was supported by the Russian Foundation for Basic Research via grant \# 13-02-91058 
(general phase diagram, Fig.2) and by the Russian Science Foundation via grant
\# 14-42-00044 (all other results, presented in Eqs.(\ref{eq:thermalcond},\ref{Lorentz},\ref{eq:modifiedpowerlaw})).

 We are grateful to  V. E. Kravtsov, K. Michaeli, L. B. Ioffe  and K. S. Tikhonov for useful discussions.

\appendix
\section{Derivation of the energy current  starting from the lattice  model.}
\label{AppendixA}
For  a general Hamiltonian which is a sum of local on-site energy operators $H = \sum_n h_{n,n+1}$, with on-site energies which satisfy continuity equation $\partial_t h_{n,n+1} + j_{E,n+1} - j_{E,n} = 0$ with energy current $j_{E,n} = i \left[h_{n-1,n},h_{n,n+1}\right]$. For particular Hamiltonian \eqref{eq:H1} on-site energies has the form:
\begin{equation}
	h_{n,n+1}  = -J\left[\hat{S}_{n}^{x} \hat{S}_{n+1}^{x} + \hat{S}_{n}^{y}S_{n+1}^{y}+\Delta \hat{S}_{n}^{z} \hat{S}_{n+1}^{z}\right]+h_{n} \hat{S}_{n}^{z}
\end{equation}
Substituting this expression into the expression for energy current yields two contributions. First contribution is of kinetic nature, it does not contain disorder and can be written in a compact form as a determinant:
\begin{equation}
	j_{E,n}^{(kin)} =\mathrm{det}\left(\begin{array}{ccc}
		\hat{S}_{n-1}^{x} & \hat{S}_{n}^{x} & \hat{S}_{n+1}^{x}\\
		\hat{S}_{n-1}^{y} & \hat{S}_{n}^{y} & \hat{S}_{n+1}^{y}\\
		\Delta\hat{S}_{n-1}^{z} & \hat{S}_{n}^{z} & \Delta\hat{S}_{n+1}^{z}
	\end{array}\right)
\end{equation}
Below we will focus only on the second term, which contains disorder. Corresponding expression in the original spin representation and in the Jordan-Wigner representation reads as follows:
\begin{eqnarray}
	j_{E,n}^{(dis)} = -i\frac{J}{2}h_{n}(S_{n-1}^{+}S_{n}^{-}-S_{n-1}^{-}S_{n}^{+}) = \nonumber \\ 
	= -i \frac{J}{2}h_{n}(c_{n-1}^{\dagger}c_{n}-c_{n}^{\dagger}c_{n-1})
\end{eqnarray}
Next step is to take continuum limit by replacing lattice operators $c_n$ with continuous field $\psi(x = n a) = c_n / \sqrt{a}$ and replacing fields $h_n$ with continuous potential $V(x = n a) = h_n$. Corresponding expression for energy current density then reads as follows:
\begin{equation}
	j_E^{(dis)}(x) = -i\frac{Ja}{2}V(x)(\psi^{\dagger}(x-a)\psi(x)-\psi^{\dagger}(x)\psi(x-a))
\end{equation}

In order to separate forward and backward scattering, we introduce slowly varying in space left- and right-moving fermionic fields $\psi_{L,R}(x)$ with $\psi(x) = e^{i k_F x} \psi_R(x) + e^{-i k_F x} \psi_L(x)$. After splitting potential $V(x)$ onto ``forward-scattering'' part $\eta(x)$ with Fourier harmonics $q \sim 0$ and ``backward-scattering'' part $\xi(x)$ with $q \sim 2 k_F$, one obtains contributions to energy current from forward- and backward-scattering processes:
\begin{eqnarray}
	& j_{E}^{(f.s.)} = \frac{J a}{2}\eta(x)(\psi_{R}^{\dagger}(x-a)\psi_{R}(x)-\psi_{L}^{\dagger}(x-a)\psi_{L}(x)) + \nonumber \\
	& + h.c. \approx Ja\eta(x)\left(\psi_{R}^{\dagger}(x)\psi_{R}(x)-\psi_{L}^{\dagger}(x)\psi_{L}(x)\right)
\end{eqnarray}
\begin{eqnarray}
	& j_{E}^{(b.s.)} = \frac{Ja}{2}\xi^{*}(\psi_{R}^{\dagger}(x-a)\psi_{L}(x)-\psi_{R}^{\dagger}(x)\psi_{L}(x-a)) + \nonumber \\
	& + h.c. \approx  \frac{Ja^{2}}{2}\xi^{*}(-\nabla\psi_{R}^{\dagger}\psi_{L}+\psi_{R}^{\dagger}(x)\nabla\psi_{L}) + h.c.
\end{eqnarray}
One can see that backward-scattering contribution is of the next order in the small  lattice constant $a$ and 
indeed vanishes in the continuum limit,  that  is $a \to 0$ keeping $u \propto J a$ constant.


\begin{thebibliography}{20}
\bibitem{BAA} D.M. Basko, I.L. Aleiner and B.L. Altshuler, Ann. Phys.\textbf{321}, 1126 (2006)
\bibitem{Mirlin05} I.V. Gornyi, A.D. Mirlin and D.G. Polyakov Phys.
Rev. Lett. \textbf{95}, 206603 (2005)
\bibitem{HuseReview} R. Nandkishore and D.A.Huse, Annual Review of
Condensed Matter Physics, \textbf{6}, 15 (2015).
\bibitem{mblpalhuse} A. Pal and D. A.Huse, Phys.Rev.B \textbf{82}, 174411 (2010)
\bibitem{Serbyn1} M. Serbyn, Z. Papić, and D. A. Abanin Phys. Rev. Lett. \textbf{111}, 127201 (2013)
\bibitem{Serbyn2} M. Serbyn, Z. Papić, and D. A. Abanin, Phys. Rev. Lett. \textbf{110}, 260601 (2013).
\bibitem{Berkovits14} R. Berkovits, Phys. Rev. B \textbf{89} 205137 (2014).
\bibitem{BarLev14} Y. Bar Lev, G. Cohen, and D.R. Reichman, Phys. Rev. Lett. \textbf{114}, 100601 (2015).

\bibitem{exp1} A.V. Sologubenko, K. Giannò, H. R. Ott, A. Vietkine, and A. Revcolevschi, Phys.Rev. B \textbf{64}, 054412 (2001).
\bibitem{exp2} N. Hlubek, P.Ribeiro, R. Saint-Martin, A. Revcolevschi, G. Roth, G. Behr, B. Büchner, and C. Hess, Phys.Rev. B 
\textbf{81}, 020405(R) (2010)
\bibitem{exp3} N. Hlubek, P. Ribeiro, R. Saint-Martin, S. Nishimoto, A. Revcolevschi, S.-L. Drechsler, G. Behr, J. Trinckauf, J. E. Hamann-Borrero, J. Geck, B. Büchner, and C. Hess
Phys. Rev. B \textbf{84}, 214419 (2011) 
\bibitem{Moore1} C. Karrasch, R. Ilan and J. E. Moore, Phys.Rev. B \textbf{88}, 195129 (2013)
\bibitem{Moore2} Y. Huang, C. Karrasch, and J. E. Moore, Phys.Rev. B \textbf{88}, 115126 (2013)
\bibitem{rggiamarchishulz} T.Giamarchi and H.J.Shulz, Phys.Rev.B, \textbf{37}, 325 (1988)
\bibitem{giamarchibook} T.Giamarchi, \textit{Quantum Physics in One Dimension} (Clarendon press, Oxford, 2003)
\bibitem{bosons2012}Z. Ristivojevic, A. Petkovic, P. Le Doussal, and T.Giamarchi
Phys. Rev. Lett. \textbf{109}, 026402 (2012)

\bibitem{1dchainNumerical} P. Schmitteckert, T. Schulze, C. Schuster, P. Schwab, and
U. Eckern, Phys. Rev. Lett. \textbf{80}, 560 (1998)
\bibitem{FIM10} M. V. Feigelman, L. B. Ioffe and M. Mezard, Phys
Rev. B \textbf{82}, 184534 (2010)
\bibitem{Klein} A. Klein and J. F. Perez, Commun. Math. Phys. \textbf{128}, 99 (1990)
\bibitem{mirlintransport} I.V.Gornyi, A.D.Mirlin, D.G.Polyakov, Phys.Rev.B, \textbf{75}, 085421
\bibitem{mirlincdw} A.D. Mirlin, D.G.Polyakov, V.M.Vinokur, Phys.Rev.Lett \textbf{99}, 156405 (2007)
\bibitem{orignac} M.-R. Li and E.Orignac, Europhys.Lett., \textbf{60}, 432-438 (2002)
\bibitem{FI15} M. V. Feigel'man and L. B. Ioffe, to be published.

\end{thebibliography}
\end{document}